\documentclass[aps,prc,twocolumn,groupedaddress,showpacs]{revtex4-1}
\usepackage{amsmath}
\usepackage{relsize}
\usepackage{graphicx}
\usepackage{textcomp}
\newcommand {\ve} [1] {\mbox{\boldmath $#1$}}
\newcommand {\beq} {\begin{eqnarray}}
\newcommand {\eol} {\nonumber \\}
\newcommand {\la} {\langle}
\newcommand {\ra} {\rangle}
\newcommand {\eeqn} [1] {\label{#1} \end{eqnarray}}%

\begin{document}
\title{Reduced sensitivity of the ($d,p$) cross sections to the  deuteron model   beyond adiabatic approximation  }
\author{M. G\'omez-Ramos$^1$ and N.K. Timofeyuk$^2$}
\affiliation{$^1$Departamento de FAMN, Universidad de Sevilla, Apartado 1065, 41080 Sevilla, Spain\\
$^2$Department of Physics, Faculty of Engineering and Physical Sciences, University of Surrey
Guildford, Surrey GU2 7XH, United Kingdom 
}

\date{\today}
\begin{abstract}
It has recently been reported   [{\it Phys. Rev. Lett. 117, 162502 (2016)}] that   $(d,p)$ cross sections can be very sensitive to the $n$-$p$ interactions 
used in the adiabatic treatment of deuteron breakup with nonlocal nucleon-target optical potentials. To understand to what extent  this sensitivity could originate in the inaccuracy of the adiabatic approximation we have developed a leading-order local-equivalent continuum-discretized coupled-channel model that 
accounts for non-adiabatic effects  in the presence of nonlocality of nucleon optical potentials. 
We have applied our model to the astrophysically relevant reaction $^{26m}$Al$(d,p)^{27}$Al  using two different 
$n$-$p$ potentials associated with the lowest and the highest  $n$-$p$ kinetic energy in the short-range region of their interaction, respectively. Our calculations   reveal a significant reduction of the sensitivity to the high $n$-$p$ momenta thus confirming that it is mostly associated with theoretical uncertainties of the adiabatic approximation itself. The non-adiabatic effects in the presence of nonlocality were found to be stronger than those in the case of the local optical potentials. These results argue for extending the  analysis of the $(d,p)$ reactions, measured for spectroscopic studies,  beyond the adiabatic approximation.
\end{abstract}

\maketitle

{\it Introduction}.  
One nucleon transfer in ($d,p$) reactions is an important source of information about the single-particle strength 
in atomic nuclei, quantified by spectroscopic factors and 
asymptotic normalization coefficients. 
They are obtained from  a comparison of experimental and theoretical cross sections calculated using direct transfer reaction theory and, therefore, are influenced by its uncertainties. The uncertaintities arising due to the input optical potentials and the shape of the mean field that binds the transferred neutron has been known for a very long time. Recently, new theoretical uncertainties have been identified in Ref. \cite{Bai16}, associated with the $n$-$p$ interaction used in adiabatic treatement of deuteron breakup with $\it nonlocal$ nucleon optical potentials. This work studied the $^{26}$Al($d,p)^{27}$Al reaction, measured in \cite{Mar15} to pin down the $^{26}$Al destruction by the $(p,\gamma)$ reactions in novae explosions, and used  several deuteron models:   Hulth\'en model \cite{Hulthen}, AV18 \cite{v18}, Reid soft core \cite{RSC}, CD-Bonn \cite{CDB} and  the chiral effective field theory at N4LO with five different regulators \cite{efts}. All these models produce exactly the same deuteron wave functions $\phi_d$ and the  vertex functions $V_{np}\phi_d$, where $V_{np}$ is the $n$-$p$ potential,  at the $n$-$p$ separations $r$ larger than than 2 fm. However, the model predictions for these quantities at $0 < r < 2$ fm are very different. This sensitivity to the short-range  $n$-$p$ wave functions (and the corresponding sensitivity to the high $n$-$p$ momenta) seems puzzling given the relatively low deuteron incoming energies, about 10 MeV, for which the ($d,p$) calculations have been done in \cite{Bai16}. Such sensitivity may indicate that other important effects, associated with ($d,p$) reaction mechanisms, are missing in these calculations.

In this paper, we  show that most of the sensitivity of the $A(d,p)B$ cross sections to the high $n$-$p$ momenta goes away when deuteron breakup is treated beyond the adiabatic distorted-wave  approximation (ADWA). The latter  is based on the dominant  term in the Weinberg state expansion of the $A$+$ n $+$ p$ wave function, calculated neglecting the couplings to all the other Weinberg components \cite{JT}. In ADWA with local $n$-$A$ and $p$-$A$ optical potentials, 
the adiabatic potential $U_{dA}(R)$, given by the sum $U_{nA}(R)+U_{pA}(R)$ \cite{JS}, does not depend on deuteron model.  However, the nonlocal adiabatic potential
explicitly depends on the average $n$-$p$ kinetic energy over the (short) range of their interaction, given by the matrix element $\la T_{np}\ra_V \equiv \la \phi_d|V_{np}T_{np}|\phi_d\ra /\la \phi_d| V_{np}|\phi_d\ra$ \cite{Tim13a,Tim13b,Bai16,Bai17}.
This matrix element is very sensitive to high $n$-$p$ momenta, which is reflected in the ADWA cross sections.


We choose the continuum-discretized coupled-channel (CDCC) approach \cite{Raw74,Aus87} to treat deuteron breakup in $A(d,p)B$ reactions beyond the adiabatic aproximation.  The CDCC,  developed and used for  local nucleon-target optical potentials only, in some cases predicts significantly different cross sections than the ADWA does
\cite{Upa12,Pan14,Cha17}. Extending the CDCC to the case of  nonlocal $n$-$A$ and $p$-$A$ potentials, in principle, could be done on the basis of the exact nonlocal ADWA formalism of Ref. \cite{Bai17}. However,  it would involve time-consuming calculations  of nonlocal kernels  when the $d$-wave component in deuteron is included, making the whole task  difficult. For this reason, based on ideas of \cite{Tim13a,Tim13b} we have developed a leading-order local-equivalent CDCC approximation    to have a quick accessment of the role of the high $n$-$p$ momenta in ($d,p$) reactions. In the ADWA, 
 the leading order solution deviates from  the exact one by about 10$\%$  but the sensitivity to the deuteron model is present  in both of them in the same proportions \cite{Bai17}, which justifies using of the leading order local-equivalent CDCC for our purposes.

{\it Nonlocal CDCC model}. 
In the CDCC, the wave function $\Phi(\ve{R},\ve{r})$ of the $A$+$n$+$p$ system includes expansion over the $n$-$p$ continuum bins $\phi_i(\ve{r})$. To begin with, we  assume that the bins represent only the $s$-wave  motion and that all spins are neglected. In this case,
\beq
\Phi(\ve{R},\ve{r}) = \sum_{i=0} \chi_i(\ve{R}) \phi_i(\ve{r})
\eeqn{wf1}
and channel function $\chi_i$ are found from the three-body nonlocal Schr\"odinger equation given by Eq. (9) of Ref. \cite{Tim13b}. In Eq. (1) and everywhere below we assume that $\phi_0$ is the deuteron bound state wave function $\phi_d$. We assume that nonlocal potentials ${\cal U}_{nA}$ and ${\cal U}_{pA}$ have the Perey-Buck form \cite{PB},
\beq
{\cal U}_{NA}(\ve{r}, \ve{r}') = H(\ve{r} -\ve{r}')U_{NA}[(\ve{r} + \ve{r}')/2],
\eeqn{PB}
with the nonlocality factor $H$ of the (small) range $\beta$,
\beq
H(x) = \pi^{-3/2}\beta^{-3}e^{-( x/\beta )^2}
\eeqn{}
and the formfactor $U_{NA}$ is given by usual Woods-Saxon form. Following the  reasoning of \cite{Tim13b} it is easy to show that  $\chi_i$ can be found from the nonlocal coupled equations
 \beq
 (T_R &+& U_C(R)- E_d) \chi_i(\ve{R}) =
 \eol
 & -& \sum_{i'} \int d\ve{s}\, H(s) {\cal V}_{ii'}(\ve{s},\ve{R}) \chi_{i'}\left(\frac{\alpha_2 \ve{s}}{2}+\ve{R}\right),
 \eeqn{nleq1}
 where $T_R$ is the kinetic energy operator, $U_C$ is the Coulomb potential energy and $E_d$ is the center-of-mass beam energy of the $d-A$ system,
 \beq
 {\cal V}_{ii'}(\ve{s},\ve{R})  = \sum_{N} \int d\ve{x}\, \phi_i^*(\ve{x}+\alpha_1\ve{s}) U_{NA}\left( \frac{\ve{x}}{2}-\ve{R}\right)\phi_{i'}(\ve{x}),  \eol \,\,\,\,\,\,\,\,\,\,
 \eeqn{}
$N$ is $n$ or $p$, $\alpha_1 = A/(A+1)$ and $\alpha_2 = (A+2)/(A+1)$. Because of the short range of $H(s)$ the wave function $\chi_{i'}\left(\frac{\alpha_2 \ve{s}}{2}+\ve{R}\right)$ can be represented by the leading-order  expansion that retains only spherical components  in $\ve{s}$   \cite{Tim13b},
\beq
 \chi_{i'}\left(\frac{\alpha_2 \ve{s}}{2}+\ve{R}\right) \approx \sum_{n=0}^{n_{\max}}
 \frac{ s^{2n}}{\beta^{2n}} 
   \gamma_n T^n_R \, \chi_{i'}(\ve{R}), \eol
  \eeqn{}
in which
  \beq
  \gamma_n = \frac{(-)^n}{n!(2n+1)!!}
 \left(\frac{\mu_d \alpha_2^2 \beta^2}{4\hbar^2} \right)^n,
 \eeqn{}
where $\mu_d$ is the reduced mass of $A+d$. Then Eqs. (\ref{nleq1}) become
 \beq
 (T_R + U_C(R) &-& E_d) \chi_i(\ve{R}) =
 \eol
 &-&
 \sum_{n=0}^{n_{\max}} \gamma_n  \sum_{i'} U^{(n)}_{ii'}(\ve{R}) T_R^n \,\chi_{i'}(\ve{R}),
 \eeqn{nleq2}
 with the coupling potentials 
 \beq
 U^{(n)}_{ii'}(\ve{R}) =  \int d\ve{x}\left[{\bar \phi}_i^{(n)}(\ve{x}) \right]^{*} \left[ \sum_{N} U_{NA}\left( \frac{\ve{x}}{2}-\ve{R}\right)\right] \phi_{i'}(\ve{x}) \eol
 \eeqn{}
 that contain the modified-by-nonlocality functions
 \beq
 \label{modbins}
 {\bar \phi}_i^{(n)}(\ve{x}) = \int d\ve{s}\, H(s) \left(\frac{s}{\beta}\right)^{2n} 
 \phi_i(\ve{x}+\alpha_1\ve{s}).
 \eeqn{}
To solve the coupled equations (\ref{nleq2}) we use the local energy-approximation.  In the case of a single channel, this approximation means $T_R \approx E - U_C - U_{loc}(R)$ with $U_{loc}$ obtained from a transcendental equation \cite{PB}. For the multichannel CDCC case we introduce a generalization of the local-energy approximation,
\beq
T_R \chi_i(\ve{R}) = \sum_k \left[(E-U_C(R))\delta_{ik} - U^{{\rm loc}}_{ik}(\ve{R})\right] \chi_k(\ve{R}). \,\,\,\,\,\,\,\,\,\,
\eeqn{CCLEA}
We apply it  $n_{\max}$ times to the r.h.s. of Eq. (\ref{nleq2}) neglecting commutators between $T_R$ and $U^{\rm loc}_{ii'}$. For one-channel case, the corrections beyond this assumption, determined by $\beta^4$,  are very small \cite{Tim13b}. Imposing the  requirement that the local-equivalent coupling potentials $U_{ii'}^{\rm loc}$ satisfy
\beq
(T_R + U_C(R) &-& E_d) \chi_i(\ve{R}) = -\sum_{i'} U_{ii'}^{\rm loc} (\ve{R}) \chi_{i'}(\ve{R}), \,\,\,\,\,\,\,\,\,
\eeqn{}
we obtain a system of the transcendental matrix equations
\beq
f_{ii'}^{(0)} - (E_j-U_C) \delta_{ij} &+& \sum_k (f_{ii'}^{(1)} +\delta_{ik})X_{ki'}
\eol 
&+& \sum_{kl}f_{ii'}^{(2)}X_{kl}X_{li'} +...=0, \,\,\,\,\,\,\,\,\,\,\,\,\,\,\,
\eeqn{nl1}
for
\beq
X_{ii'} = (E_{i'}-U_C) \delta_{ii'} - U^{\rm loc}_{ii'},
\eeqn{nlineq1}
in which $f_{ij}^{(n)}= \gamma_n  U^{(n)}_{ii'}$.
We solve equations (\ref{nl1}) using Newton method and then read $U_{ii'}^{\rm loc}$  into the CDCC reaction code, which in our case was FRESCO \cite{fresco}.

The scheme described above remains unchanged when all  spins are included. We will assume in the following that the target has spin 0, although it can be proved that a non-zero target spin simply introduces an overall factor in the coupling potentials. In the coupling scheme, consistent with FRESCO ($\ve{l}+\ve{s_n} = \ve{j_n}$, $\ve{j_n}+\ve{s_p}=\ve{I}$ and $\ve{L}+\ve{I}=\ve{J}$),
 the bin functions $\phi_{\alpha}$ are labeled by a set of quantum numbers $\alpha = \{i,l, j_n\}$, where $i$ includes both the bin energy and its total angular momentum $I$, $l$ is the $n$-$p$ orbital momentum and $j_n$ is the total momentum of neutron.
The channel functions $\chi_{iLJ}$ depend on the  $d-A$ relative orbital momentum $L$ and total momentum $J$. We require that the local-equivalent coupling potentials $ U_{ii'\lambda}^{\rm loc} $  satisfy
\beq
(T_R&+& U_C(R) - E_d) \chi_{iLJ}(R) = 
\eol
&-&\sum_{i'L'\lambda} {\cal C}^{LIJ}_{I'L'\lambda} \,U_{ii'\lambda}^{\rm loc} (R) \,\chi_{i'L'J}(R), \,\,\,\,\,\,\,\,\,\,\,\,\,\,\,\,
\eeqn{}
where the (un)primed quantities correspond to the (initial) final state,
\beq
{\cal C}^{LIJ}_{I'L'\lambda} = (-)^{I+L+J}
\left\{ 
\begin{array}{ccc}
 L & I & J \\ 
 I' & L' & \lambda
\end{array}
\right\}
\hat{L}' \la L'0 \lambda 0 |L0\ra, \,\,\,\,\,\,\,\,\,\,\,\,\,\,\,\,
\eeqn{}
\begin{widetext} and $\hat{a} = \sqrt{2a+1}$.  
 The $ U_{ii'\lambda}^{\rm loc} $  are also found from a system of  transcendental matrix equations 
\beq
g_{ii'\lambda}^{(0)} &-& (E_{i'}-U_C) \hat{I}\delta_{ii'} \delta_{\lambda0} 
+
\sum_{k_1l_1 l_2j_1} (g_{ik_1l_1}^{(1)}+\hat{I}\delta_{ik_1} \delta_{l_1 0} ) \dfrac{\hat{\lambda}}{\hat{l_2}}{\cal C}^{\lambda II'}_{j_1 l_2 l_1}X_{k_1i'}^{(l_2)}
+\sum_{
\begin{array} {c}
{\scriptstyle   k_1k_2j_1j_2 }\\ 
{\scriptstyle  l_1l_2l_3\Lambda_1} 
\end{array}}
g_{ik_1l_1}^{(2)}\dfrac{\hat{\Lambda}_1}{\hat{l_2}} 
{\cal C}^{\Lambda_1 Ij_2}_{j_1 l_2 l_1}{\cal D}^{\lambda II'}_{j_2l_3 \Lambda_1}
X_{k_1k_2}^{(l_2)}X_{k_2i'}^{(l_3)}
\eol
&+&
\sum_{
\begin{array} {c}
{\scriptstyle  k_1k_2k_3 j_1j_2j_3 } \\ 
{\scriptstyle  l_1l_2l_3l_4 \Lambda_1\Lambda_2}
\end{array}}
g_{ik_1l_1}^{(3)} \,
\dfrac{\hat{\Lambda}_1}{\hat{l_2}} 
{\cal C}^{\Lambda_1 Ij_2}_{j_1 l_2 l_1} \,
{\cal D}^{\Lambda_2Ij_3}_{j_2l_3 \Lambda_1 }\,
{\cal D}^{\lambda II'}_{j_3l_4 \Lambda_2}
X_{k_1k_2}^{(l_2)}X_{k_2k_3}^{(l_3)}X_{k_3i'}^{(l_4)}
\,+...=0, \,\,\,\,\,\,\,\,\,\,\,\,\,\,\,
\eeqn{nl2}
\end{widetext}
with ${\cal D}^{LIJ}_{I'L'\lambda}=(-)^{I'-I}\hat{L}\hat{L'}^{-1}{\cal C}^{LIJ}_{I'L'\lambda}$ and $j_i$ being the spin of state $k_i$,
written for 
\beq
X_{ii'}^{(\lambda)} = (E_{i'}  - U_C)\hat{I}' \delta_{ii'}\delta_{\lambda 0} - U^{\rm loc}_{ii'\lambda}.
\eeqn{}
Eqs. (\ref{nl2}) now include all necessary angular momentum couplings.
They contain functions
\begin{equation}
\begin{split}
g_{ii'\lambda}^{(n)}(R)&= \gamma_n \sum_{ll'j_nj'_n} (-)^{l+s_n+s_p+j_n+j_n'+I'}\hat{I}\hat{I}'\hat{j}_n\hat{j}_n'\frac{\hat{\lambda}^2\hat{l}}{4\pi} \\ &\times 
\left\lbrace 
\begin{array}{ccc}
 j_n & j_n' & \lambda \\ 
 I' & I & s_p
\end{array}
\right\rbrace
\left\lbrace 
\begin{array}{ccc}
 j_n & j_n' & \lambda \\ 
 l' & l & s_n
\end{array}
\right\}U^{(n)}_{\alpha\alpha'\lambda}(R),
\end{split}
\end{equation}
determined by the multipoles of the coupling potentials folded between the original $\phi_i$ and modified ${\bar \phi}_i$ functions:
\begin{equation}
\begin{split}
U^{(n)}_{\alpha\alpha'\lambda}(R) =&
 \int_0^\infty dx \, x^2\left[{\bar \phi}_{\alpha}^{(n)}(x) \right]^{*} \left[ \sum_N U^{(\lambda)}_{NA}\left(x,R\right)\right] \phi_{\alpha'}(x)\\
U^{(\lambda)}_{NA}\left(x,R\right) &=
2\pi \int_{-1}^1 du \,\,U_{NA}\left(\frac{\ve{x}}{2}-\ve{R}\right) P_\lambda (u),
\end{split}
\end{equation}
with $u$ being the cosine between $\ve{x}$ and $\ve{R}$ and $P_\lambda (u)$ the Legendre polynomial.

{\it  Application to the $^{26m} {\rm Al}(d,p)^{27} {\rm Al}$ reaction}. We apply the newly developed local-equivalent CDCC model to the $(d,p)$ reaction recently  measured in inverse kinematics with isomeric $^{26m}$Al  beam  
\cite{Alm17}. Because of the 0$^+$ spin of this isomer 
transfers to the final $^{27}$Al states will involve only one orbital momentum, thus facilitating extraction of spectroscopic factors.

We have performed the CDCC calculations for three incident deuteron energies, 9.2, 25 and 50 MeV, typical for the TRIUMF, GANIL and RIKEN facilities. We used the Gianinni-Ricco systematics of energy-independent nonlocal nucleon optical potentials for $N=Z$ targets \cite{GR} and two nucleon-nucleon (NN)  potentials: Hulth\'en and  RSC. In Ref. \cite{Bai16} the calculations with these potentials gave the lowest and the highest $^{26g}$Al($d,p)^{27}$Al cross sections, respectively. Both $s$- and $d$-wave continuum bins were used in the calculations. For the reaction at 9.2 MeV three bins were taken for each component considered equispaced for proton-neutron energies from 0 to 6 MeV (closed channels start at 6.3 MeV). At 25 MeV, five bins were taken from 0 to 20 MeV (closed channels at 20.97 MeV) and at 50 MeV, four bins from 0 to 44 (closed channels at 44.18 MeV). Convergence with bin mesh was checked in calculations with local potentials  at all energies and with nonlocal potentials at 9.2 MeV. We were also made aware that contributions from the closed channels at low $E_d$ are  negligible \cite{Rub18}.


\begin{figure}[b]
\includegraphics[width=0.5\textwidth]{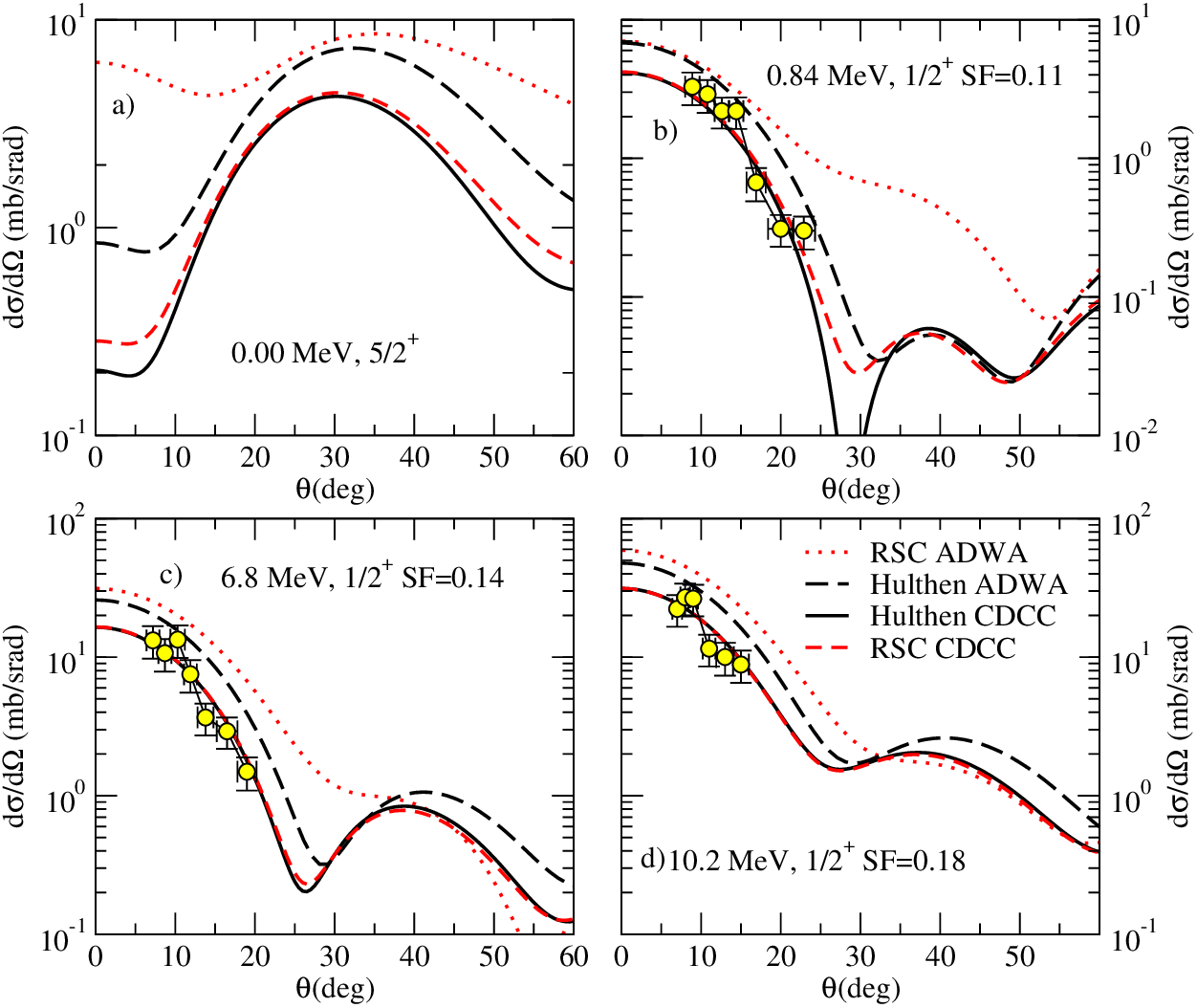}
\caption{
The  differential cross sections of $^{26m}$Al($d,p)^{27}$Al at $E_d^{\rm lab} = 9.2$ MeV for population of the $^{27}$Al($5/2^+$) ground state (a) and the excited $1/2^+$ states at $E_x = 0.84$, 6.8 and 10.2 MeV ($b,c,d$), respectively. In the cases with experimental data \citep{Alm17} all calculations have been multiplied by the spectroscopic factor obtained for the RSC-CDCC calculation.
}
\label{fig:1}
\end{figure}

\begin{table*}[t]
\caption{Various ratios of the $^{26m}$Al($d,p)^{27}$Al cross sections calculated in the ADWA and CDCC   with two different NN potentials, RSC and Hulth\'en, for $E_d^{\rm lab} = 9.2$, 25 and 50 MeV and for four final states in $^{27}$Al. The ratios were calculated at the maxima of cross secitons. All energies are given in MeV.
}
\centering
\begin{tabular} {p {1.8 cm}  p {1.  cm} p{ 1. cm} p{ 1.5 cm} p{ 1. cm} p{ 1. cm} p{ 1.5 cm} p{ 1. cm} p{ 1. cm} p{ 1.5 cm} p{ 1. cm} p{ 1. cm} p{ 0.6 cm} }
\hline\hline
 &\multicolumn{3}{c}{$\sigma^{\rm ADWA}_{\rm RSC}/\sigma^{\rm ADWA}_{\rm H}$ \,\,\,\,\,\,\,\,\,\,\,\,\,\,\,\,\,} 
 & \multicolumn{3}{c}{$\sigma^{\rm CDCC}_{\rm RSC}/\sigma^{\rm CDCC}_{\rm H}$ \,\,\,\,\,\,\,\,\,\,\,\,\,\,\,\,\,}
 & \multicolumn{3}{c}{$\sigma^{\rm ADWA}_{\rm H}/\sigma^{\rm CDCC}_{\rm H}$ \,\,\,\,\,\,\,\,\,\,\,\,\,\,\,\,\,}
 & \multicolumn{3}{c}{$\sigma^{\rm ADWA}_{\rm RSC}/\sigma^{\rm CDCC}_{\rm RSC}$ \,\,\,\,\,\,\,\,\,\,\,\,\,\,\,\,\,}
 \\
 \cline{2-13}  
$E_x$\textbackslash$ E_d^{\rm lab}$ & 9.2 & 25 & 50 & 9.2 & 25 & 50 & 9.2 & 25 & 50 & 9.2 & 25 & 50 \\
\hline
0.00 &  1.17 & 1.38 & 1.76 & 1.04 & 1.08 & 1.05 & 1.71 & 1.80 & 1.55 & 1.81 & 2.35 & 2.60 \\
0.84 &  1.03 & 1.12 & 2.24 & 1.01 & 0.97 & 1.14 & 1.64 & 1.29 & 1.78 & 1.67 & 1.50 & 3.54 \\
6.8  &  1.21 & 0.86 & 2.09 & 1.00 & 0.96 & 1.07 & 1.57 & 1.28 & 1.37 & 1.89 & 1.14 & 2.68 \\
10.2 &  1.24 & 0.81 & 1.69 & 1.00 & 0.96 & 1.05 & 1.52 & 1.27 & 1.10 & 1.89 & 1.08 & 1.76 \\
\hline\hline
\end{tabular}
\label{tab:ratios}
\end{table*}

We have calculated the local-equivalent coupling potentials $U^{\rm loc}_{ij\lambda}(R)$ at each point $R$ from 0 to 50 fm by solving Eq. (\ref{nl2})  using the Newton method. The choice of  $n_{\max} = 3$ was sufficient for $U^{\rm loc}_{ij\lambda}$ to converge, similar to findings in the one-channel study  \cite{Tim13b}. For some $d$-wave channels, $n_{\max}=2$ was sufficient.  The  $U^{\rm loc}_{ij\lambda}$ have been read  into FRESCO which calculated the channel functions $\chi_i$ and then
the finite range transfer cross sections using the same NN potentials in the transfer vertex. In the case of the RSC,  both the $s$- and $d$-wave deuteron vertex  functions were used. The $\la^{26m}$Al$|^{27}$Al$\ra$ overlap function was represented by the neutron single-particle wave function, calculated for  the Woods-Saxon potential well with the standard  radius $r_0=$ 1.25 fm and diffusseness $a$ = 0.65 fm.


\begin{figure}[b]
\includegraphics[width=0.5\textwidth]{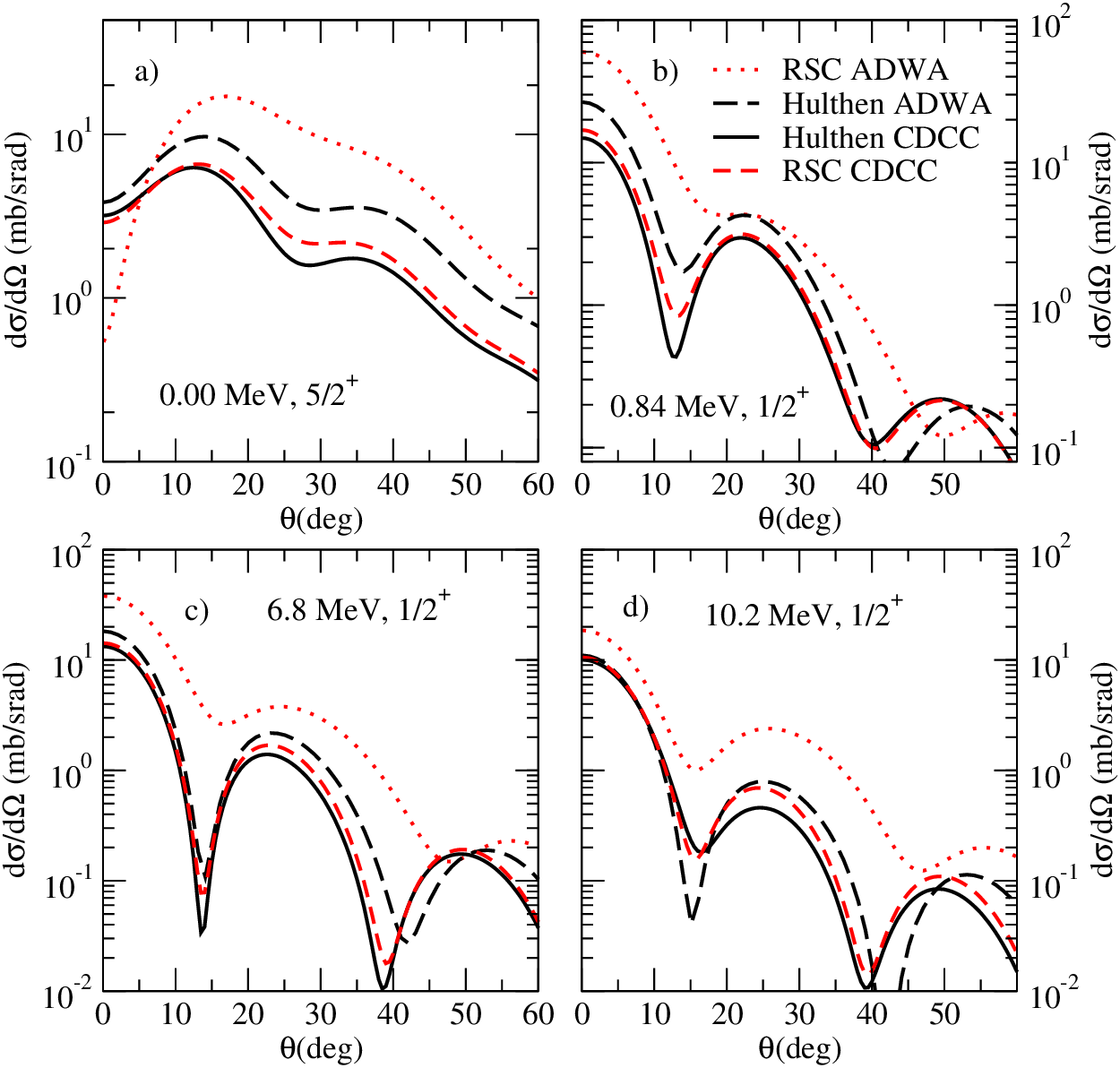}
\caption{
The same as in Fig. 1 but for $E_d^{\rm lab} = 50$ MeV.
}
\label{fig:2}
\end{figure}

The leading order nonlocal CDCC and ADWA calculations are shown in Fig. 1 and 2 for deuteron incident energies of 9.2 and 50 MeV, respectively, and for four final $^{27}$Al states: the ground $J^{\pi}=5/2^+$ state  and three astrophysically revelant excited $J^{\pi}=1/2^+$ states. In all cases, the CDCC cross sections are significantly lower than the ADWA ones. Their ratio in the maximum, shown in Table I, in most cases is higher than an average value of 1.25 reported for local optical potentials in \cite{Cha17}. 
The ratio seems to correlate with the neutron separation energy in the final state:  for $l=0$ transfer to the final $1/2^+$ state it decreases with excitation energy.


The ADWA cross sections, obtained with Hulth\'en and RSC, differ up to a factor of two in the maximum (see Table I), which is related to the  small and large values of the matrix element $\la T_{np}\ra_V$ associated with these potentials \cite{Bai16}.  
The Hulth\'en-ADWA calculations effectively include only $s$-wave continuum while RSC-ADWA includes the $d$-wave continuum as well. To check to what extent the difference between these calculations is due to the missing $s$-wave continuum we performed the Hulth\'en-CDCC calculations with $s$-wave bins only for one selected case, $J^{\pi}=5/2^+$ at $E_d^{\rm lab} = 9.2$ MeV. The cross sections were 10$\%$ lower thus pointing that the $n$-$p$ model dependence in ADWA partially originates from a different $d$-wave content of continuum associated with these models.

The CDCC calculations show that the sensitivity to the $n$-$p$ model is significantly reduced. It is less than 4$\%$ for $E_d^{\rm lab} = 9.2 $ MeV but can understandably increase  with the deuteron incident energy up to 14$\%$. 

Although our main aim is the comparison of ADWA and CDCC calculations, given the existence of experimental data for  $^{27}$Al($1/2^+$) \citep{Alm17}, we deduced spectroscopic factors from these data using  both ADWA and CDCC and both  NN potentials. They are presented in Table \ref{tab:sf} and  compared to previous ADWA calculations with local optical potentials.
Both CDCC calculations and the Hulth\'en-ADWA reproduce the shape of experimental data but RSC-ADWA  overestimates the data at larger angles for the states at $E_x=$ 0.84 and 6.8 MeV. The spectroscopic factors extracted with CDCC are larger than those determined in \citep{Alm17}, but this difference decreases with the excitation energy.


\begin{table}[b]
\caption{Spectroscopic factors obtained from the $^{26m}$Al($d,p)^{27}$Al($1/2^+$) cross sections calculated with ADWA and CDCC with two different NN potentials, RSC and Hulth\'en for $E_d^{\rm lab} = 9.2$ MeV. All energies are in MeV.
}
\centering
\begin{tabular} {p {1.8 cm} p{ 1.2 cm} p{ 1.2 cm} p{ 1.2 cm} p{ 1.2 cm} p{ 1. cm} }  
$E_x$ & ADWA Hulth\'en & ADWA RSC & CDCC Hulth\'en & CDCC RSC & Ref \citep{Alm17}\\
\hline
\hline
0.84 &  0.07 & -- & 0.13 & 0.11 & 0.08\\
6.8  &  0.14 & -- & 0.14 & 0.14 & 0.11\\
10.2 &  0.13 & 0.08 & 0.18 & 0.18 & 0.16\\
\hline\hline
\end{tabular}
\label{tab:sf}
\end{table}

\begin{figure}[t]
\includegraphics[width=0.45\textwidth]{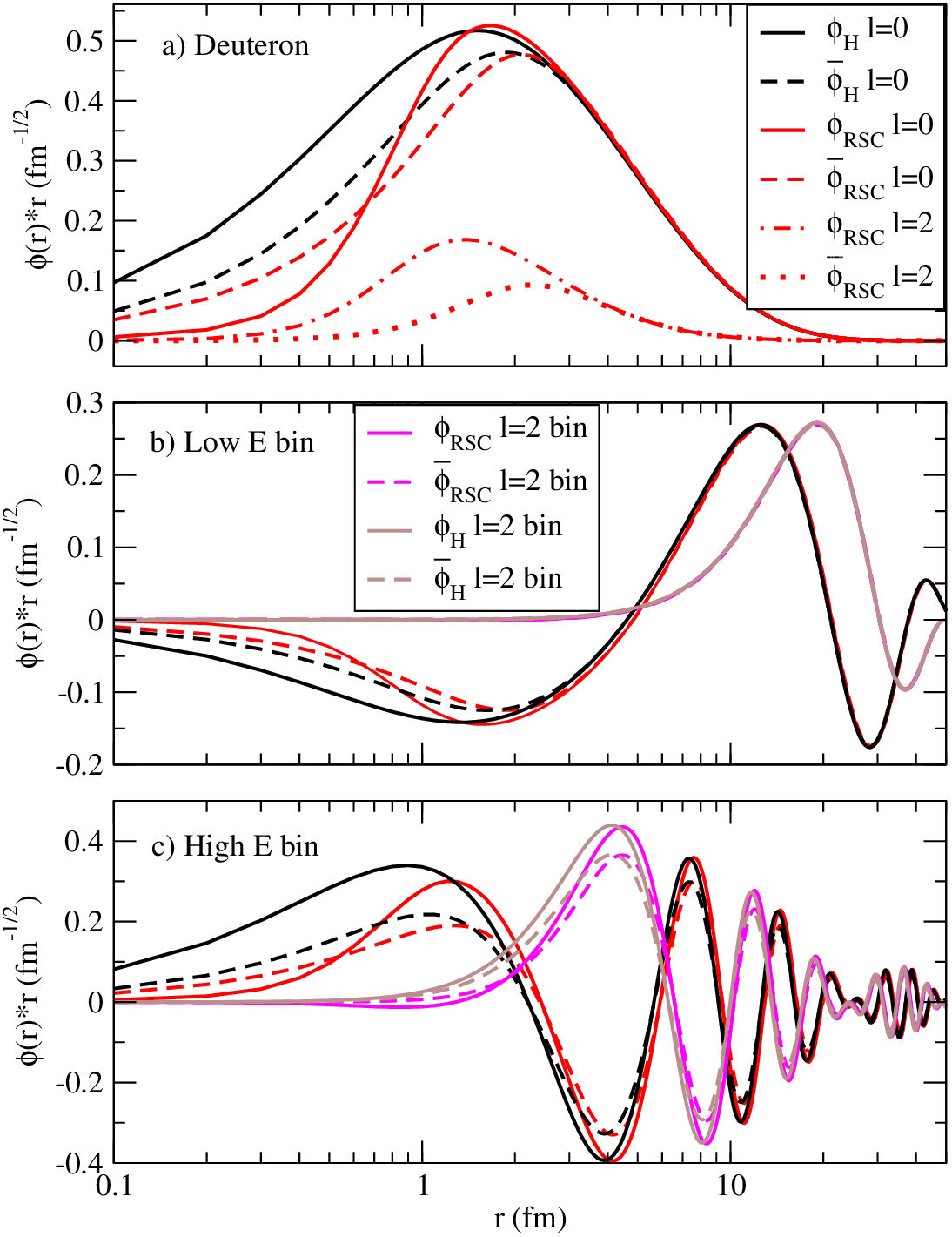}
\caption{
Original $\phi_i$ and modified-by-nonlocality ${\bar \phi}_i$ deuteron ($a$) and $J^{\pi}=1^+$ bin ($b,c$) wave functions obtained with RSC and Hulth\'en potentials. The low-energy bin ($b$) has an everage energy $\bar{E}=1.6$ MeV with a spread of $\Delta E=2$ MeV while the high-energy bin  ($c$) corresponds to $\bar{E}=34.375$ MeV and $\Delta E=19.25$ MeV. For both energies, the bins with incoming $s$ and $d$ waves are presented. 
}
\label{fig:3}
\end{figure}

{\it Understanding reduced sensitivity}. The strong sensitivity of the ADWA cross sections to the NN model comes from the coefficient $M_0$ in the transcendental equation  for the local-equivalent adiabatic potential $U^{\rm loc}$,
\beq
U^{\rm loc} = M_0 (U_{nA}+U_{pA})\exp \left[-\gamma (E - U_C - U^{\rm loc})\right],\,\,\,\,\,\,\,\,\,\,\,\,
\eeqn{ulocad}
where $\gamma$ is a constant \cite{Tim13b}. This coefficient is given by
\beq
M_0 &=& N\int d\ve{s} d\ve{x} \,H(s)  \phi_d^*(\ve{x}+\alpha_1 \ve{s})V_{np}(\ve{x})\phi_d(\ve{x})
 \,\,\,\,\,\,\,\, 
\eeqn{M0}
with $N = \la \phi_d | V_{np}| \phi_d\ra^{-1}$ (see \cite{Tim13b} for the link between $M_0$ and $\la T_{np}\ra_{V}$). Because of the short range of $V_{np} \phi_d$, $M_0$ is highly sensitive to the  details of $\phi_d$ at  small $x$. In the CDCC, the  main channel corresponds to the folding model with the $U^{\rm loc}$   found from Eq. (\ref{ulocad})  and $M_0$ generated by Eq. (\ref{M0}) with $N=1$ and without $V_{np}$ \cite{JSnl}:
\beq
M_0 = \int  d\ve{x} \,  {\bar \phi}^*_d(\ve{x})\phi_d(\ve{x}).
 \,\,\,\,\,\,\,\, 
\eeqn{M0fold}
Because of the small deuteron binding energy this $M_0$ is determined by the large values of $x$, corresponding to small $n$-$p$ momenta,  where all the NN models agree. Also, because of the small range of nonlocality $\beta$, in this range  ${\bar \phi}_d \approx \phi_d$ (see Fig. \ref{fig:3}a) and, therefore, $M_0 \approx 1$. The same statements are relevant for low-energy continuum bins which are affected by the nonlocality and differences in the NN potentials only at small $x$ ( Fig. \ref{fig:3}b)   thus explaining the reduced sensitivity to the deuteron model in the ($d,p$) calculations with CDCC. The differences in the NN model affect high-energy bins ( Fig. \ref{fig:3}c) where modifications due to nonlocality are stronger. As a result, the sensitivity to the high $n$-$p$ momenta is stronger for a large deuteron incident energy, as seen from Table I.


The ADWA could be corrected  by including more   Weiberg states in the  expansion of $\Phi(\ve{R},\ve{r})$    \cite{JT,Laid}. This would involve calculations of  nondiagonal local-equivalent coupling potentials $U^{\rm loc}_{ii'}$ that depend on the coefficients given by  (\ref{M0}) but with Weinberg states $\phi^W_i$ instead of $\phi_d$. Such coefficients (and, therefore, the $U^{\rm loc}_{ii'}$ and the corresponding ($d,p$) cross sections)  would be determined by the model-dependent short-range behaviour of $V_{np}\phi^W_i$. It was shown in \cite{Pan13} that continuum bins could be expanded over Weinberg states. Therefore, sufficient number of NN-dependent Weinberg states should recover the almost-NN-independent CDCC calculations. It is worth mentioning that for local optical potentials the non-adiabatic corrections  explicitly depend  on the same NN model-dependent matrix element $\la T_{np}\ra_V$  \cite{Joh14} that features in the nonlocal ADWA.

{\it Conclusions}. Based on our newly developed local-equivalent CDCC model with nonlocal optical potentials, we have shown that the previously reported strong sensitivity of the adiabatic $(d,p)$ cross sections, calculated with $nonlocal$ nucleon optical potentials, is significantly reduced. 
For low deuteron incident energies it is now less than 4$\%$ but can increase up to 14$\%$ for higher energies.


We have also found that  non-adiabatic effects are much stronger than those in the case of local optical potentials. To confirm this finding, the nonlocal CDCC should be extended beyond the leading order. Exact ADWA cross sections with nonlocal potentials are smaller than the leading-order cross sections \cite{Bai17} but  this tendency may not necessarily be the same in the CDCC case. It is conceivable that the difference between exact nonlocal CDCC and nonlocal ADWA can be smaller than that obtained in this work.
 
The  sensitivity to high $n$-$p$ momenta  due to uncertainties of the adiabatic approximation suggests that theoretical analysis of  $(d,p)$ experiments should be extended beyond the adiabatic approximation when nonlocal optical potentials are used. This is an important message given the current interest of other groups in ADWA with nonlocal potentials, such as in \cite{Tit16,Ros16}. Full nonlocal CDCC calculations could help to refine the spectroscopic factors and asymptotic normalization coefficients obtained from $(d,p)$ reactions. We note that present results were obtained with energy-independent optical potentials. A proper treatment of energy-dependence within the three-body context is a challenge, in particular in the CDCC formalism, where the energy between nucleon and target is not well defined in the considered final states. Whether approximate prescriptions to take this dependence into account could result in additional NN-model dependence of (d,p) cross sections remains to be investigated.

{\it Acknowledgements}.
We are grateful  to R. C. Johnson and A. M. Moro for fruitful discussions. This work was supported by the United Kingdom Science and Technology Facilities Council (STFC) under Grant No. ST/L005743/1. M. G.-R. acknowledges a research grant from the Spanish Ministerio de Educaci\'on, Cultura y Deporte, Ref: FPU13/04109.

\bibliographystyle{apsrev4-1}

\end{document}